\documentstyle[epsfig,aps,pre,multicol]{revtex}
\begin{document}
\draft
\title{Continuity Conditions for the Radial Distribution Function\\
of Square-Well Fluids}
\author{L.\ Acedo}
\address{Departamento de F\'{\i}sica, Universidad  de  Extremadura,\\
E-06071 Badajoz, Spain}
\date{\today}
\maketitle
\begin{abstract}
The continuity properties of the radial distribution function $g(r)$
and its close relative the cavity function $y(r) \equiv e^{\phi(r)/k_B T} g(r)$
are studied in the context of the Percus-Yevick (PY) integral equation for 3D
square-well fluids. The cases corresponding to a well width, $(\lambda-1)
\sigma$, equal to a fraction of the diameter of the hard core, $\sigma/m$, with
$m=1,2,3$ have been considered. In these cases, it is proved that
the function $y(r)$ and its first derivative are everywhere continuous but
eventually the derivative of some order becomes discontinuous at the points
$(n+1) \sigma/m$, $n=0,1,\ldots$. The order of continuity (the highest order
derivative of $y(r)$ being continuous at a given point), $\kappa_n$, is found
to be $\kappa_n \sim n$ in the first case ($m=1$) and $\kappa_n \sim 2 n$ in the
other two cases ($m=2,3$), for $n \gg 1$. Moreover, derivatives of $y(r)$ up
to third order are continuous at $r=\sigma$ and $r=\lambda \sigma$ for
$\lambda=3/2$ and $\lambda=4/3$ but only the first derivative is continuous
for $\lambda=2$. This can be understood as a non-linear resonance effect.
\end{abstract}
\section{INTRODUCTION}
\label{sect_1}
Simple models have played an important role in the development of
liquid state theory, both as approximations and as useful reference
systems in perturbation schemes. The simplest model
with a repulsive hard core and an attractive well is the
square-well (SW) fluid. The SW interaction potential is
\begin{equation}
\label{SWI}
\phi(r)=\left\{ \begin{array}{rl} \infty , & \quad r < \sigma \\
\noalign{\smallskip}
-\epsilon, & \quad \sigma < r < \lambda \sigma \\
\noalign{\smallskip}
0, & \quad r > \lambda \sigma ,
\end{array} \right.
\end{equation}
where $\sigma$ is the diameter of the hard core, $\epsilon$ is the well
depth, and $(\lambda-1) \sigma$ is the well width. The equilibrium properties
of the fluid depend on the values of three dimensionless parameters: the fraction
of volume occupied by the spheres $\eta = \pi \rho \sigma^3 / 6$ ($\rho$
being the number density), the reduced
temperature $T^*=k_B T / \epsilon$ ($T$ being the temperature and $k_B$
being the Boltzmann constant), and the width parameter $\lambda$. These
properties can be derived from a more fundamental quantity, the so-called
radial distribution function $g(r)$ \cite{MTY,Rev}. The quantity
$4 \pi r^2 g(r) dr / V  $ is the probability that 
the centers of two spherical atoms in the liquid are separated by 
a distance between $r$ and $r+d r$, $V$ being the volume of the system. In 
the absence of particle interactions (ideal gases) there is no structure 
in the fluid and $g(r)=1$. This is also true for any realistic potential in the limit 
$r \rightarrow \infty$ as we must have $\phi(r) \rightarrow 0$ in that limit.
It is also evident that the centers of two particles in the SW fluid cannot
be nearer than $\sigma$ due to the hard core, so $g(r)=0$ for $r < \sigma$.\\
Many closed integral equations have been proposed for this function
(YBG, HNC, Percus-Yevick, \ldots), but the most popular is, perhaps, the
Percus-Yevick (PY) equation. In the search for these approximations it has
been found useful to define an auxiliary function, the direct
correlation function $c(r)$, through the Ornstein-Zernike (OZ) relationship
\begin{equation}
\label{OZ}
h(\vert {\bf r}_2-{\bf r}_1 \vert)=c(\vert {\bf r}_2-{\bf r}_1 \vert)+
\rho \int\, d^3 {\bf r}_3 c(\vert {\bf r}_1-{\bf r}_3 \vert) h(\vert
{\bf r}_3-{\bf r}_2 \vert) 
\end{equation}
where $h(r)=g(r)-1$. In this equation and the following ones we are
assuming that the interaction potential is spherically symmetric and the
structure functions, consequently, depends only on the distance between
the particle centers. The physical meaning of the direct correlation
function is clear; it accounts only for the correlation effects arising
from a direct interaction between the particles in the fluid. Nevertheless,
the total correlation function, $h(r)$, comprises also the correlations
propagated by intermediate particles. The direct correlation function is
expressed in terms of the so-called bridge functionals, which constitute
an infinite sum of diagrams \cite{MTY}. Integral equations are obtained
by closing the OZ relation with an approximation for $c(r)$ based upon
a sum of some set of these diagrams. By choosing an appropiate set of
diagrams, Percus and Yevick\cite{PY,Percus} showed that the direct correlation
function is approximately given by
\begin{equation}
\label{cpy}
c(r)=g(r) \left[ 1-e^{\phi(r)/k_B T} \right]
\end{equation}
According to (\ref{cpy}) the range of $c(r)$ is equal to that of the
interaction potential $\phi(r)$ as $c(r)$ is zero in those regions
where the interaction potential vanishes. The PY equation is given
by the substitution of (\ref{cpy}) into the OZ relation (\ref{OZ})
and it takes the form \cite{MTY}
\begin{equation}
\label{PYMTY}
y(\vert {\bf r}_2 - {\bf r}_1 \vert)=A+ \rho \int \, d^3  {\bf r}_3
 \, f(\vert {\bf r}_1-{\bf r}_3 \vert)
\left[1+f(\vert {\bf r}_3 - {\bf r}_2 \vert) \right] y(\vert {\bf r}_1
-{\bf r}_3 \vert) y(\vert {\bf r}_3 - {\bf r}_2 \vert) 
\end{equation}
where $y(r) \equiv e^{\phi(r)/k_B T} g(r)$ is the cavity function, $f(r)=e^{-\phi(r)/k_B T}-1$
is the Mayer function and $A$ is a constant defined as
\begin{equation}
\label{Adef}
A=1-\rho \int_0^\infty \, dr f(r) y(r) 4 \pi r^2 
\end{equation}
The PY equation is exactly solvable for the hard-sphere (HS) fluid
\cite{WT} and the sticky-hard-sphere (SHS) fluid\cite{BX}. Both models can 
be considered as special cases of the more general SW interaction: the HS 
potential is recovered if we take the limit $\epsilon \rightarrow 0$ (i.e., 
$T^* \rightarrow \infty$) or $\lambda \rightarrow 1$; the SHS fluid is defined by 
taking the limits $\lambda \rightarrow 1$ and $\epsilon \rightarrow \infty$
(i.e., $T^* \rightarrow 0$) simultaneously, while keeping the parameter
$\tau^{-1}=12 \left( 1 - \lambda^{-1} \right) e^{1/T^*}$ constant.\\ 
Nevertheless, neither the PY equation nor any other integral equation for fluids
has ever been analytically solved for the SW interaction. In 1977, Sharma and Sharma
\cite{Sharma} proposed a mean spherical approximation which provides 
an analytical expression for the structure factor but it is not consistent 
with the hard core exclusion constraint. By the same time, Nezbeda 
\cite{Nezbeda} proposed a polynomial approximate solution for $y(r)$ valid in the 
limit $\lambda-1 \ll 1$. This solution was based on the continuity of the
first and the second derivatives of $y(r)$ at $r=\sigma$. More recently, Yuste 
and Santos\cite{YusSan} derived a simple approximate expression for the Laplace 
transform of $r g(r)$ for the SW fluid. This derivation is based upon simple
physical requirements (finiteness of the radial distribution function at
$r=\sigma^{+}$ and finite isothermal compressibility) and the continuity of $y(r)$ at
$r=\lambda$.\\
The aim of this paper is the analysis of the continuity properties of
the function $y(r)$ satisfying the PY equation for the
three-dimensional SW fluid. First, we must identify the points
at the borders of the regions where $y(r)$ is an analytic function. For 
$\lambda=2$ it is clear that these points are given by $r=\sigma,
2\sigma,\ldots$, as also happens for the HS potential \cite{MTY}. The
radial distribution function is then divided in analytic pieces $\psi_0(r)$,
$\psi_1(r)$, $\ldots$ and the PY equation (\ref{PYMTY}) is written as a
system of non-linear integral equations for them. In these equations
the Heaviside step function that enters in Eq. (\ref{PYMTY})
through the Mayer function no longer appears. The continuity properties are
derived from this system. The cases $\lambda=3/2$ and $\lambda=4/3$ are more
cumbersome as the intervals cited above are divided into two and
three equal parts, respectively, and the number of
equations is obviously larger. More general cases with $1 < \lambda < 4/3$ are
increasingly more difficult to manage since the number of intervals becomes
larger and larger as the well width, $\lambda-1$, becomes smaller. We
will not deal with those cases in this paper.\\
The paper is organized as follows. In Sec.\ \ref{sect_2} the PY
equation is set in a form that fits better our purpose and the continuity
of $y(r)$ and its first derivative for $r=\lambda \sigma$ is suggested
from the simulation results of Henderson {\em et al.}\cite{Henderson}.
The appropiate system of integral equations for the case
$\lambda=2$ is written in Sec.\ \ref{sect_3}; the general derivatives are
then related with the derivatives of lower order and the continuity conditions
are derived recursively. A similar procedure is used in Sec.\ \ref{sect_4}
for the cases $\lambda=3/2$ and $\lambda=4/3$. The conditions for the HS
fluid are given in an Appendix. The paper ends with some remarks on
the relevance of these conditions for the proposal of approximations.
\section{DEFINITIONS AND BASIC EQUATIONS}
\label{sect_2}
In the applications of the PY equation to simple potentials is
often useful to define a new structure function $\psi(r)=r y(r)$, which
satisfies the same continuity conditions as $y(r)$. After the change of
variables $t=\vert {\bf r}_2-{\bf r}_1\vert/\sigma$, $x=\vert {\bf r}_1-{\bf 
r}_3 \vert/\sigma$ and $y=\vert {\bf r}_3-{\bf r}_2\vert/\sigma$ one can 
easily check that Eq. (\ref{PYMTY})
takes the form \cite{MTY}
\begin{equation}
\label{PYPSI}
\psi(t)= A t+12 \eta \int_0^{\infty}\, d x f(x) \psi(x)\, \int_{\vert t
-x \vert}^{t+x}\, d y \left[1+f(y)\right] \psi(y) 
\end{equation}
where 
\begin{equation}
\label{APSI}
A=1-24\eta\int_0^\infty\, d x f(x) x \psi(x)
\end{equation}
It is remarkable that this reduction is only possible in three dimensions
(excepting the one-dimensional case, where no reduction is necessary) and that
the same change of variables in two dimensions yields a significatively more
complex integral equation due to the lack of rotational simmetry around the
vector ${\bf r}_2-{\bf r}_1$. The Mayer function $f(x)=\mbox{exp}\left[-\phi(x)/k_B T
\right]-1$ for the SW interaction (\ref{SWI}) is given by
\begin{equation}
\label{fx}
1+f(x)=H(x-1)\left[ 1+\alpha H(\lambda-x) \right]
\end{equation}
where $H(x)$ is the Heaviside step function and 
$\alpha\equiv\mbox{exp}\left(\epsilon/k_B T\right)-1$. The first derivative of 
$\psi(t)$ obtained from (\ref{PYPSI}) is also given here for future reference
\begin{figure}
\begin{center}
\epsfig{file=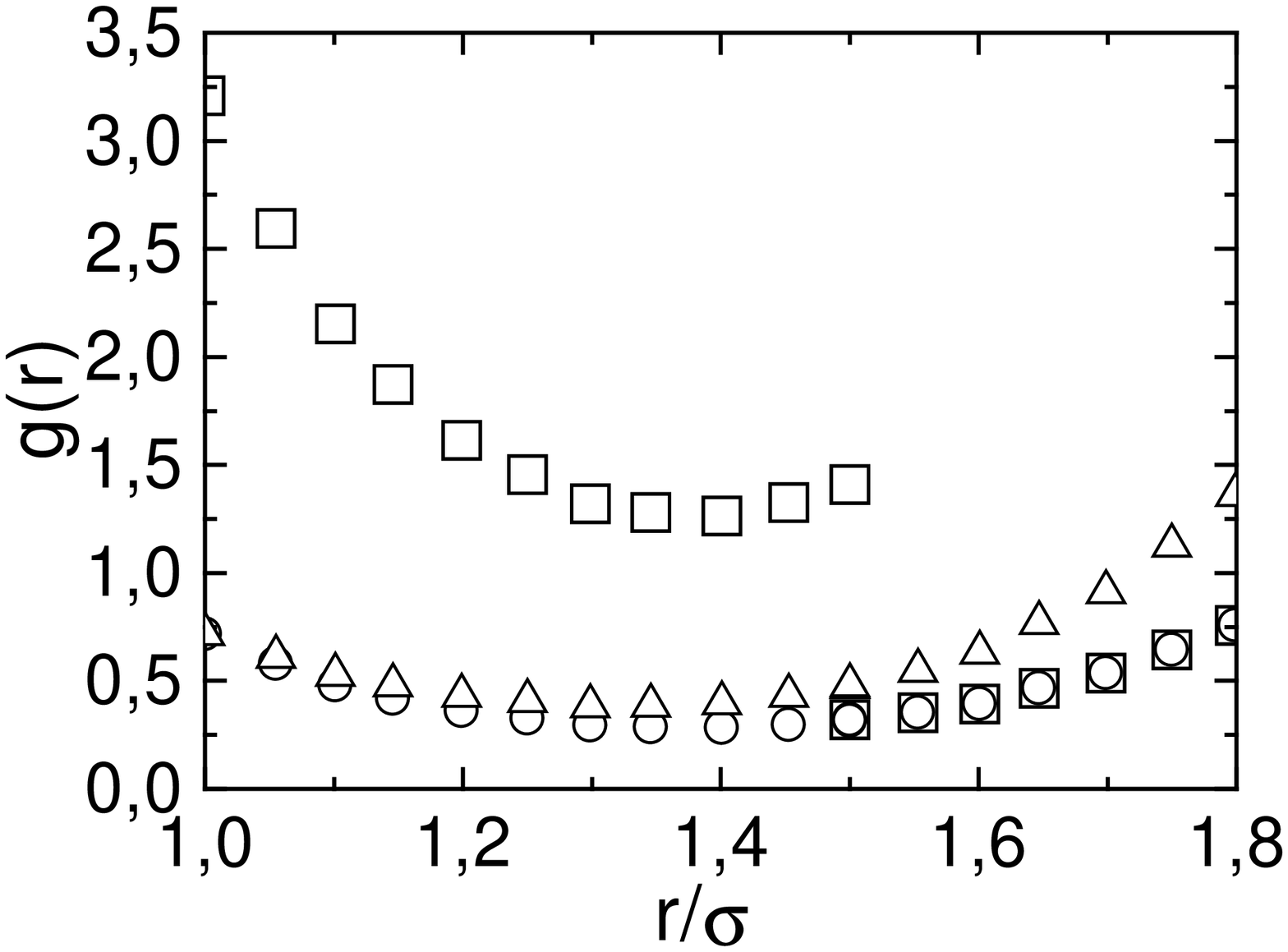,width=4in,clip=}
\caption{Monte Carlo simulation results\protect\cite{Henderson} for the 
structure functions $g(r)$ (squares), $y(r)$ (circles), and 
$\psi(r)$ (triangles) on a SW fluid with $\lambda=3/2$, $T^*=2/3$
and $\eta=2\pi/15$. Note that $g(r)$ and $y(r)$ overlap
for $r > \lambda \sigma$.\label{fig1}}
\end{center}
\end{figure}
\begin{equation}
\label{PYFD}
\begin{array}{rcl}
\psi^{'}(t)&=&A+12 \eta \displaystyle\int_0^\infty\, d x f(x) \psi(x) 
\left[1+f(x+t)\right]
\psi(x+t) \\
\noalign{\smallskip}
& &-12 \eta \displaystyle\int_0^t\, d x f(x) \psi(x) \left[1+f(t-x)\right] 
\psi(t-x) \\
\noalign{\smallskip}
& &+12 \eta \displaystyle\int_t^\infty\, d x f(x) \psi(x) \left[ 1+f(x-t) 
\right] \psi(x-t) 
\end{array}
\end{equation}
In Fig. \ref{fig1} the simulation results of Henderson 
{\em et al.}\cite{Henderson} for $g(r)$, $y(r)$ and $\psi(r)$ for
a SW fluid with $\lambda=3/2$, $T^*=2/3$ and $\eta=2\pi/15$ are 
shown. The radial
distribution function is not continuous at $r=\lambda \sigma$ but both
$y(r)$ and $\psi(r)$ are clearly continuous functions at that point. 
Simulation results also suggest the continuity of the first derivative but
nothing can be said about the second or higher order derivatives as these
are very difficult to measure in a simulation, where only results for
widely spaced values of $r$ are provided.
\section{THE SQUARE-WELL FLUID WITH $\lambda=2$}
\label{sect_3}
In this case the well width coincides with the diameter of the hard core
and the function $\psi(t)$ that satisfies the corresponding 
PY equation (\ref{PYPSI}) is conveniently defined 
in a piecewise fashion as follows
\begin{equation}
\label{PSIPL}
\psi(t)=\left\{\begin{array}{lr}
\psi_0(r) & \quad 0 < r < 1 \\
\noalign{\smallskip}
\psi_1(r) & \quad 1 < r < 2 \\
\noalign{\smallskip}
\vdots & \quad \vdots \\
\noalign{\smallskip}
\psi_n(r) & \quad n < r < n+1 \\
\noalign{\smallskip} 
\vdots & \quad \vdots  
\end{array} \right.
\end{equation}
where we assume, to be confirmed later, that $\psi_n(r)$, $n=0,1,\ldots$
are analytic functions in their intervals of definition. This
will be justified by expressing (\ref{PYFD}) as a system of integral
equations for these functions where the Heaviside step function
in (\ref{fx}) no longer appears. Thus, any derivative of $\psi_n(r)$, 
$n=0,1,\ldots$ is related to derivatives of lower order of the same 
set of functions. It can be expected that at a given 
point $t=n+1$, $n=0,1,\ldots$ the functions $\psi_n(r)$ and $\psi_{n+1}(r)$
do not match perfectly and this gives rise to a difference between the
derivatives of some order of these functions evaluated at that point. 
We define the symbols
\begin{equation}
\label{defi}
\Delta_n^{(k)}=\psi_n^{(k)}(n+1)-\psi_{n+1}^{(k)}(n+1)  \quad 
n=0,1,\ldots, \quad  k=0,1,\ldots 
\end{equation}
as the jump in the derivative of order $k$ evaluated at $t=n+1$. The
continuity of $\psi(t)$ is a well known fact \cite{Nezbeda,YusSan}, so that
we have $\Delta_n^{(0)}=0$, $n=0,1,\ldots$. The first derivative is
given in (\ref{PYFD}) in terms of integrals over the functions $\psi(t)$
and $f(t)$, which in turn depends on the Heaviside step function. Starting 
from (\ref{PYFD}) and (\ref{fx}) the following set of equations
for $\psi_n^{'}(t)$, $n=0,1,\ldots$ is found
\begin{eqnarray}
\psi_0^{'}(t)&=&A-12 \eta (\alpha+1) \int_{1-t}^1\,
d x \psi_0(x) \psi_1(x+t) \nonumber \\
\noalign{\smallskip}
& &+12 \eta \alpha (\alpha+1) \left\{\int_{1+t}^2\, d x \psi_1(x) 
\psi_1(x-t)+\int_1^{2-t} \, d x \psi_1(x) \psi_1(x+t) \right\} \nonumber \\
\noalign{\smallskip}
\label{1a}
& &+12 \eta \alpha \int_{2-t}^2 \, d x \psi_1(x) \psi_2(x+t) 
\end{eqnarray}
\begin{eqnarray}
\psi_1^{'}(t)&=&A-12 \eta \left\{ \int_0^{2-t} \, d x \psi_0(x) \psi_1(x+t)
\right. \nonumber \\
\noalign{\smallskip}
 & &\left. + \int_{2-t}^1 \, d x \psi_0(x) \psi_2(x+t)-\int_0^{t-1} \, d x
\psi_0(x) \psi_1(t-x) \right\} \nonumber \\
\noalign{\smallskip}
 & &+12 \eta \alpha\left\{ \int_1^{3-t} \, d x \psi_1(x) \psi_2(x+t)+
\int_{3-t}^2 \, d x \psi_1(x) \psi_3(x+t) \right. \nonumber \\
\noalign{\smallskip}
\label{1b}
 & &\left.-\int_0^{2-t}\, d x \psi_0(x) \psi_1(x+t)+\int_0^{t-1}\, d x
\psi_0(x) \psi_1(t-x) \right\} 
\end{eqnarray}
\begin{eqnarray}
\psi_2^{'}(t)&=&A-12 \eta \left\{ \int_0^{3-t}\, d x \psi_0(x) \psi_2(x+t)
+\int_{3-t}^1 \, d x \psi_0(x) \psi_3(x+t) \right. \nonumber \\
\noalign{\smallskip}
& &\left. -\int_0^{t-2}\, dx \psi_0(x) \psi_2(t-x)-\int_{t-2}^1\, d x 
\psi_0(x)
\psi_1(t-x) \right\} \nonumber \\
\noalign{\smallskip}
& &+12 \eta \alpha \left\{ \int_1^{4-t}\, d x \psi_1(x) \psi_3(x+t) +
\int_{4-t}^2\, d x \psi_1(x) \psi_4(x+t) \right. \nonumber \\
\noalign{\smallskip}
& &\left. -\int_1^{t-1} \, d x \psi_1(x) \psi_1(t-x)+\int_{t-2}^1\, d x
\psi_0(x) \psi_1(t-x) \right\} \nonumber \\
\noalign{\smallskip}
\label{1c}
& &-12 \eta \alpha^2 \int_1^{t-1} \, d x \psi_1(x) \psi_1(t-x)  
\end{eqnarray}
\begin{eqnarray}
\psi_n^{'}(t)&=&A-12\eta\left\{\int_0^{n+1-t}\, d x \psi_0(x) \psi_n(x+t)
+\int_{n+1-t}^1\, d x \psi_0(x) \psi_{n+1}(x+t) \right. \nonumber \\
\noalign{\smallskip}
& &\left. -\int_0^{t-n}\, d x \psi_0(x) \psi_n(t-x) - \int_{t-n}^1 \, d x
\psi_0(x) \psi_{n-1}(t-x) \right\} \nonumber \\
\noalign{\smallskip}
& &+12 \eta \alpha \left\{ \int_1^{n+2-t} \, d x \psi_1(x) \psi_{n+1}(x+t)
+\int_{n+2-t}^2 \, d x \psi_1(x) \psi_{n+2}(x+t) \right. \nonumber \\
\noalign{\smallskip}
& &\left.-\int_1^{t-n+1} \, d x \psi_1(x) \psi_{n-1}(t-x)-\int_{t-n+1}^2 \, 
d x \psi_1(x) \psi_{n-2}(t-x) \right\} \nonumber \\
\noalign{\smallskip}
\label{1d}
& &-12 \eta \alpha^2 \int_{t-2}^2 \, d x \psi_1(x) \psi_1(t-x) \delta_{n,3}\, ,
\quad n \ge 3 
\end{eqnarray}
With the PY equation written in the form of this system it is
easy to show that $\Delta_n^{(1)}=0$, $n=0,1,\ldots$ as a consequence of 
the cancelation of some of the integral terms at the border points.
The continuity of $\psi(t)$ and its first derivative is also true in the
cases $\lambda=3/2$, $\lambda=4/3$, and in the HS fluid and it is possibly
a general property of the PY equation. The derivatives of order
$k$, $k \ge 2$ of $\psi_1(t)$, $\psi_2(t)$, etc\ldots are obtained from
the system of equations (\ref{1a})--(\ref{1d}), yielding
\begin{eqnarray}
\psi_0^{(k)}(t)&=&-12 \eta \sum_{j=0}^{k-2}\, (-1)^j \left\{ (\alpha+1) 
\psi_0^{(j)}(1-t) \psi_1^{(k-2-j)}(1)+\alpha (\alpha+1) 
\psi_1^{(k-2-j)}(1+t) \psi_1^{(j)}(1) \right. \nonumber \\
\noalign{\smallskip}
\label{2a}
& &\left. +\alpha (\alpha+1) \psi_1^{(j)}(2-t) \psi_1^{(k-2-j)}(2)
-\alpha \psi_1^{(j)}(2-t) \psi_2^{(k-2-j)}(2) \right\}
+\left[\mbox{Int}\right] \; , \\
\noalign{\smallskip}
\psi_1^{(k)}(t)&=& 12 \eta \sum_{j=0}^{k-2}\, (-1)^j \left\{ 
\psi_0^{(j)}(2-t) \Delta_1^{(k-2-j)}+(1+\alpha) (-1)^j \psi_0^{(j)}(t-1) 
\psi_1^{(k-2-j)}(1) \right. \nonumber \\
\noalign{\smallskip}
\label{2b}
& &\left.-\alpha \psi_1^{(j)}(3-t) \Delta_2^{(k-2-j)}
+\alpha \psi_0^{(j)}(2-t) \psi_1^{(k-2-j)}(2) \right\}
+\left[\mbox{Int}\right]  \\
\noalign{\smallskip}
\psi_2^{(k)}(t)&=&12 \eta \sum_{j=0}^{k-2} \left\{ (-1)^j \psi_0^{(j)}(3-t) 
\Delta_2^{(k-2-j)}-\psi_0^{(j)}(t-2) \Delta_1^{(k-2-j)}-\alpha (-1)^j
\psi_1^{(j)}(4-t) \Delta_3^{(k-2-j)} \right. \nonumber \\
\noalign{\smallskip}
\label{2c}
& &\left. -\alpha\psi_0^{(j)}(t-2) \psi_1^{(k-2-j)}(2)-\alpha(\alpha+1)
\psi_1^{(j)}(t-1) \psi_1^{(k-2-j)}(1) \right\}+\left[\mbox{Int}\right] \\
\noalign{\smallskip}
\psi_n^{(k)}(t)&=&12 \eta \sum_{j=0}^{k-2} \left\{ (-1)^j 
\psi_0^{(j)}(n+1-t) \Delta_n^{(k-2-j)}-\psi_0^{(j)}(t-n) 
\Delta_{n-1}^{(k-2-j)} \right. \nonumber \\
\noalign{\smallskip}
\label{2d}
& &\left.-\alpha (-1)^j \psi_1^{(j)}(n+2-t) \Delta_{n+1}^{(k-2-j)}+
\alpha \psi_1^{(j)}(t+1-n) \Delta_{n-2}^{(k-2-j)} \right\} \nonumber \\
\noalign{\smallskip}
& &+12 \eta \alpha^2 \delta_{n,3} \sum_{j=0}^{k-2} \psi_1^{(j)}(t-2)
\psi_1^{(k-2-j)}(2)+\left[ \mbox{Int} \right]\, , \quad n \ge 3 
\end{eqnarray}
where the terms denoted by $\left[ \mbox{Int} \right]$ include the sum
of several integrals over the functions $\psi_n(t)$, $n=0,1,\ldots$ and
their derivatives. These terms are always continuous and their explicit
expressions are not required for the calculation of the derivative jumps, so
they will not be quoted here. If we take into account the continuity of $\psi(t)$
and its first derivative, as well as the condition $\psi(0)=0$, the first
nonzero derivative jump at the border points $t=1,2,\ldots$ is
readily derived from Eqs. (\ref{2a})--(\ref{2d}):
\begin{eqnarray}
\label{3a}
\Delta_0^{(2)}&=&\Delta_2^{(2)}=-24 \eta \alpha (\alpha+1) \psi(1) \psi(2) \\
\noalign{\smallskip}
\label{3b}
\Delta_1^{(2)}&=&12 \eta (\alpha+1)^2 \left[ \psi(1) \right]^2 \\
\noalign{\smallskip}
\label{3c}
\Delta_3^{(2)}&=&12 \eta \alpha^2 \left[ \psi(2) \right]^2 \\
\noalign{\smallskip}
\label{3d}
\Delta_{2 n}^{(2 n)}&=& -12 \eta (\alpha+1) \psi(1)
\Delta_{2 n-1}^{(2 n-2)}+12 \eta \alpha \psi(2) \Delta_{2 n -2}^{(2 n -2)}\; ,
\quad n=2,3,\ldots \\
\noalign{\smallskip}
\label{3e}
\Delta_{2 n+1}^{(2 n)}&=& 12 \eta \alpha \psi(2)
\Delta_{2 n -1}^{(2 n -2)}\; , \quad n=2,3,\ldots 
\end{eqnarray}
Therefore, the order of continuity (i.e., the highest order derivative of
$\psi(r)$ being continuous) is $\kappa_n=1$ for $n \le 3$, $\kappa_n=n-1$
for $n=4,6,\ldots$, and $\kappa_n=n-2$ for $n=5,7,\ldots$. The presence
of discontinuities in the potential at $r=\sigma$ and $r=2\sigma$ gives rise
to discontinuities in the second derivative of the structure function
$\psi(r)$ not only at those points but also at their neighbors
$r=3\sigma,4\sigma$ and, surprisingly, these
discontinuities are of the same order as those at the hard core and
square-well borders. At larger distances the structure function becomes
more and more continuous and the order of continuity grows linearly. The
derivative jumps in (\ref{3d}) and (\ref{3e}) may, indeed, increase
exponentially at certain physical conditions but this is balanced with
the factorial in a Taylor expansion. These results are then compatible with
the asymptotic limit $g(r)=y(r) \rightarrow 1$, or equivalently $\psi(r)
\rightarrow r$, as $r \rightarrow \infty$. It can also be noticed that the
first nonzero derivative jump in Eqs. (\ref{3a})--(\ref{3e}) depends only
on the packing fraction $\eta$, the temperature (through $\alpha$), and
the values of the function $\psi(r)$ at the potential discontinuity points.
The order of continuity of the function $\psi(r)$ and the cavity function
coincide by definition. The derivative jumps of the second derivative of
$y(r)$ at $r=\sigma$ and $r=2 \sigma$ are given by
\begin{eqnarray}
\left. \frac{d^2 y}{d r^2} \right\vert_{r=\sigma^{+}}-
\left. \frac{d^2 y}{d r^2} \right\vert_{r=\sigma^{-}}&=&
\frac{48 \eta}{\sigma^2} \alpha (\alpha+1) y(\sigma) y(2 \sigma) \\
\noalign{\smallskip}
\left. \frac{d^2 y}{d r^2} \right\vert_{r=2\sigma^{+}}-
\left. \frac{d^2 y}{d r^2} \right\vert_{r=2\sigma^{-}}&=&
-\frac{6 \eta}{\sigma^2} (\alpha+1)^2 \left[y(\sigma)\right]^2
\end{eqnarray}
which are a consequence of Eqs. (\ref{3a}) and (\ref{3b}).
\section{THE SQUARE-WELL FLUIDS WITH $\lambda=3/2$ AND $\lambda=4/3$}
\label{sect_4}
In the case $\lambda=3/2$ the function $\psi(t)$ is also piecewise but
we must distinguish between the left and the right halves of the intervals
$n < t < n+1$, $n=0,1,\ldots$:
\begin{equation}
\label{PSIPLB}
\psi(t)=\left\{ \begin{array}{lr}
\psi_0(t) & \quad 0 < t < \frac{1}{2} \\
\noalign{\smallskip}
\overline{\psi}_0(t) & \quad \frac{1}{2} < t < 1 \\
\noalign{\smallskip}
\vdots & \quad \vdots \\
\noalign{\smallskip}
\psi_n(t) & \quad n < t < n+\frac{1}{2} \\
\noalign{\smallskip}
\overline{\psi}_n(t) & \quad n+\frac{1}{2} < t < n+1 \\
\noalign{\smallskip} 
\vdots & \quad \vdots  
\end{array} \right.
\end{equation}
Consequently, the derivative jumps at the points $t=n+1/2$ and
$t=n+1$, $n=0,1,2,\ldots$ are denoted by two sets of symbols,
$\widetilde{\Delta}_n^{(k)}$ and $\Delta_n^{(k)}$, respectively. These symbols are
defined as follows
\begin{eqnarray}
\label{LD}
\widetilde{\Delta}_n^{(k)}&=&\psi_n^{(k)}(n+\frac{1}{2})
-\overline{\psi}_n^{(k)}(n+\frac{1}{2}) \\
\noalign{\smallskip}
\label{LDb}
\Delta_n^{(k)}&=&\overline{\psi}_n^{(k)}(n+1)-\psi_{n+1}^{(k)}(n+1)  
\end{eqnarray}
As in the previous case the functions $\psi(t)$ and its first derivative
are everywhere continuous and these symbols take the value zero for
$k=0$ and $k=1$, $n=0,1,\ldots$. The equations for $\psi_n^{(k)}(t)$ and $\overline{\psi}_n^{(k)}$
are obtained from (\ref{fx}) and (\ref{PYFD}) after the elimination of the Heaviside
functions. A straightforward but lengthy calculation that runs in parallel to
that of Sec. \ref{sect_3} leads to the following results for the first
nonzero symbols at every border point
\begin{eqnarray}
\label{6a}
\Delta_0^{(4)}&=&(12 \eta)^2 (1+\alpha) \psi(1) \left\{ (1+\alpha)^2
\left[ \psi(1) \right]^2+2 \alpha^2 \left[ \psi(3/2) \right]^2
\right\} \\
\noalign{\smallskip}
\label{6b}
\Delta_1^{(2)}&=&12 \eta (1+\alpha)^2 \left[ \psi(1) \right]^2  \\
\noalign{\smallskip}
\label{6c}
\Delta_2^{(2)}&=&12 \eta \alpha^2 \left[ \psi(3/2) \right]^2 \\
\noalign{\smallskip}
\label{6d}
\Delta_n^{(2n-2)}&=&-12\eta \left\{ \psi(1) \Delta_{n-1}^{(2 n-4)}-
\alpha \psi(3/2) \widetilde{\Delta}_{n-1}^{(2 n -4)}\right\}\; , \quad n\ge 3  \\
\noalign{\smallskip}
\label{6e}
\widetilde{\Delta}_0^{(2)}&=&\widetilde{\Delta}_2^{(2)}=-24 \eta \alpha (1+\alpha)
\psi(1) \psi(3/2)   \\
\noalign{\smallskip}
\label{6f}
\widetilde{\Delta}_1^{(4)}&=&-(12 \eta)^2 \alpha \psi(3/2) \left\{ 
\alpha^2
\left[\psi(3/2)\right]^2+2 (1+\alpha)^2 \left[\psi(1)\right]^2
\right\} \\
\noalign{\smallskip}
\label{6g}
\widetilde{\Delta}_n^{(2n-2)}&=&-12 \eta \left\{(1+\alpha) \psi(1)
\widetilde{\Delta}_{n-1}^{(2n-4)}
-\alpha \psi(3/2) \Delta_{n-1}^{(2 n -4)}\right\}\; , \quad n \ge 3  
\end{eqnarray}
In this case we will denote the order of continuity at $t=n+1$ by
$\kappa_n$, $n=0,1,\ldots$ and the corresponding one at $t=n+1/2$
by $\overline{\kappa}_n$, $n=0,1,\ldots$. According to (\ref{6a})--(\ref{6f})
we conclude that $\kappa_0=\overline{\kappa}_1=3$, $\kappa_1=\kappa_2=
\overline{\kappa}_0=\overline{\kappa}_2=1$, and $\kappa_n=\overline{\kappa}_n=2 n -3$ for
$n \ge 3$. It is a remarkable fact that the cavity function $y(r)$ is
continuous up to the third order derivative at those points where the interaction
potential exhibits discontinuities, $r=\sigma$ and $r=3\sigma/2$. Nevertheless,
only the first derivative is continuous at the neighbour
points $r=\sigma/2$, $r=2\sigma$, $r=5\sigma/2$ and $r=3\sigma$. The situation is
rather different in the case studied in the previous Section. There, the
discontinuity on the square-well is located precisely at a point where
the hard core induces a discontinuity jump on the structure function. This
special disposition of the potential jumps reinforces the discontinuity of
the structure functions, $y(r)$ and $\psi(r)$, at the lattice sites $r/\sigma=1,2,\ldots$
on a kind of ``resonance'' effect. This is also observed at large distances
from the hard core as the highest order derivative exhibiting no jumps grows
as $r$ in the SW fluid with $\lambda=2$ and as $2 r$ in the SW fluid
with $\lambda=3/2$, where the resonance condition is broken.\\
In the case $\lambda=4/3$ we can find discontinuities at the points $t=n+1/3$,
$t=n+2/3$, and $t=n+1$, $n=0,1,2,\ldots$. In a similar way to the definitions
in (\ref{defi}), (\ref{LD}), and (\ref{LDb}) the symbols
$\widetilde{\Delta}_n^{(k)}$, $\widehat{\Delta}_n^{(k)}$ and
$\Delta_n^{(k)}$ are introduced as follows
\begin{eqnarray}
\widetilde{\Delta}_n^{(k)}&=&\lim_{\epsilon \to 0} \left[
\left. \frac{d^k \psi(t)}{d t^k}\right\vert_{t=n+1/3-\epsilon}
-\left. \frac{d^k \psi(t)}{d t^k}\right\vert_{t=n+1/3+\epsilon} \right] 
\\
\noalign{\smallskip}
\widehat{\Delta}_n^{(k)}&=&\lim_{\epsilon \to 0} \left[
\left. \frac{d^k \psi(t)}{d t^k}\right\vert_{t=n+2/3-\epsilon}
-\left. \frac{d^k \psi(t)}{d t^k}\right\vert_{t=n+2/3+\epsilon} \right]  \\
\noalign{\smallskip}
\Delta_n^{(k)}&=&\lim_{\epsilon \to 0} \left[
\left. \frac{d^k \psi(t)}{d t^k}\right\vert_{t=n+1-\epsilon}
-\left. \frac{d^k \psi(t)}{d t^k}\right\vert_{t=n+1+\epsilon} \right]
\end{eqnarray}
where $n=0,1,\ldots$, $k=0,1,\ldots$. The calculation of these symbols
is completely analogous to that of the previous cases save for the existence
of three independent analytic functions on the intervals $(n,n+1)$, $n=0,1,\ldots$
and the details will not be given. The results for the nonzero symbols
with the lowest value of $k$ for $n=0,1,2$ are
\begin{eqnarray}
\label{7a}
\widetilde{\Delta}_0^{(2)}&=&\widetilde{\Delta}_2^{(2)}=
-24 \eta \alpha (1+\alpha) \psi(1) \psi(4/3)\\
\noalign{\smallskip}
\label{7b}
\widehat{\Delta}_0^{(4)}&=&-(12 \eta)^2 \alpha (1+\alpha)^2
\left[\psi(1)\right]^2 \psi(4/3)  \\
\noalign{\smallskip}
\label{7c}
\Delta_0^{(4)}&=&(12\eta)^2 (1+\alpha) \psi(1) \left\{(1+\alpha)^2 \left[
\psi(1) \right]^2+2 \alpha^2 \left[\psi(4/3)\right]^2 \right\} \\
\noalign{\smallskip}
\label{7d}
\widetilde{\Delta}_1^{(4)}&=&-(12 \eta)^2 \alpha \psi(4/3)
\left\{\alpha^2\left[\psi(4/3)\right]^2+2(1+\alpha)^2 \left[
\psi(1) \right]^2\right\} \\
\noalign{\smallskip}
\label{7e}
\widehat{\Delta}_1^{(4)}&=&(12 \eta)^2 \alpha^2 (1+\alpha) \psi(1) \left[
\psi(4/3)\right]^2  \\
\noalign{\smallskip}
\label{7f}
\Delta_1^{(2)}&=&12 \eta (1+\alpha)^2 \left[ \psi(1) \right]^2  \\
\noalign{\smallskip}
\label{7g}
\widehat{\Delta}_2^{(2)}&=&12 \eta \alpha^2
\left[ \psi(4/3) \right]^2 \\
\noalign{\smallskip}
\label{7h}
\Delta_2^{(4)}&=&-12 \eta (1+\alpha) \Delta_1^{(2)}
\end{eqnarray}
Between these conditions and (\ref{6a})--(\ref{6g}) corresponding to
the case $\lambda=3/2$ we notice some similarities. In particular, the
order of continuity at the points $t=\displaystyle\frac{1}{2},1,\displaystyle\frac{3}{2},2,\displaystyle\frac{5}{2}$ and $3$
for $\lambda=3/2$ is the same that the order of continuity at the points
$t=\displaystyle\frac{1}{3},1,\displaystyle\frac{4}{3},2,\displaystyle\frac{7}{3}$ 
and $\displaystyle\frac{8}{3}$ for $\lambda=4/3$. Moreover, the
first nonzero derivative jump has the same form in both cases if we
write them in terms of $\psi(1)$ and $\psi(\lambda)$ (See Eqs. (\ref{6e})
and (\ref{7a}), (\ref{6a}) and (\ref{7c}), etc\ldots). These considerations
suffice us to conjecture that the order of continuity at the points
$t=\lambda-1,2,\lambda+1,2\lambda$ is $k=1$ and the order of continuity
at $t=1,\lambda$ is $k=3$ for any value of $\lambda$ in the interval
$1 < \lambda < 2$. The special ``resonant'' case $\lambda=2$ is excluded
from this rule. The first nonzero derivative jumps at these points in
terms of the cavity function are expected to be given by
\begin{eqnarray}
\label{8a}
\left. \frac{d^2 y}{d r^2}\right\vert_{(\lambda \pm 1)\sigma^+}-
\left. \frac{d^2 y}{d r^2}\right\vert_{(\lambda \pm 1)\sigma^-}&=&
\frac{24 \eta \lambda}{(\lambda \pm 1)\sigma^2} \alpha(1+\alpha) y(\sigma)
y(\lambda\sigma) \\
\noalign{\smallskip}
\label{8b}
\left. \frac{d^4 y}{d r^4}\right\vert_{\sigma^+}-
\left. \frac{d^4 y}{d r^4}\right\vert_{\sigma^-}&=&
-\frac{(12 \eta)^2}{\sigma^4} (1+\alpha) y(\sigma)\left\{\left[ (1+\alpha)
y(\sigma) \right]^2+2 \left[\alpha \lambda y(\lambda\sigma) 
\right]^2\right\} \\
\noalign{\smallskip}
\label{8c}
\left. \frac{d^4 y}{d r^4}\right\vert_{\lambda\sigma^+}-
\left. \frac{d^4 y}{d r^4}\right\vert_{\lambda\sigma^-}&=&
\frac{(12 \eta)^2}{\sigma^4} \alpha y(\lambda\sigma) \left\{
\left[ \alpha \lambda y(\lambda\sigma) \right]^2+2 \left[ (1+\alpha) y(\sigma) \
\right]^2 \right\} \\
\noalign{\smallskip}
\label{8d}
\left. \frac{d^2 y}{d r^2}\right\vert_{2\sigma^+}-
\left. \frac{d^2 y}{d r^2}\right\vert_{2\sigma^-}&=&
-\frac{6 \eta}{\sigma^2} \left[ (1+\alpha) y(\sigma) \right]^2 \\
\noalign{\smallskip}
\label{8e}
\left. \frac{d^2 y}{d r^2}\right\vert_{2\lambda\sigma^+}-
\left. \frac{d^2 y}{d r^2}\right\vert_{2\lambda\sigma^-}&=&
-\frac{6\eta\lambda}{\sigma^2} \left[ \alpha y(\lambda\sigma) \right]^2
\end{eqnarray}
which are a consequence of the generalization of (\ref{6a})--(\ref{6c}),
(\ref{6e}) and (\ref{6f}) or the corresponding conditions for $\lambda=4/3$.
Results for the order of continuity versus $r/\sigma$ for the 
SW fluids we have analyzed and the hard-sphere case discussed
in the Appendix are summarized in Fig.\ \ref{fig2}. The main feature of these
series of values is their linear (HS) or staircase (SW with $\lambda=2$, $3/2$ and
$4/3$) behaviour beyond a given distance which depends upon the well width. This
is a simple mathematical consequence of the explicit convolution nature of the
PY equation. It can be shown that a solution of this kind of equations with
order of continuity $p$ at $r=a$ and $p^{'}$ at $r=b$ is necessarily a 
function with order of continuity $p+p^{'}+3$ at $r=a+b$. This rule is 
sufficient to explain the trend for $r \ge 3 \sigma$ on every case plotted in
Fig.\ \ref{fig2}. Discontinuities of the structure functions derivatives
appear at $r=\sigma$ and $r=\lambda \sigma$ as a consequence of
the singularities of the interaction potential, $\phi(r)$, and are then 
propagated in
a way we are not familiar with in the domain of linear physics. The cases
analyzed in detail in this paper give us also some insight on the interplay of 
the relative positions of the discontinuites of the potential and we 
could suggest an explanation for the qualitatively different behaviours of
the cases $ 1 < \lambda < 2$ and
$\lambda=2$ showed in Fig.\ \ref{fig2}. The order of continuity appears to 
be generally lower in the
latter case and we can attribute that to an interference of the singularities of the
Mayer function because for $\lambda=2$ the square-well border is placed 
precisely on a point
\begin{figure}
\begin{center}
\epsfig{file=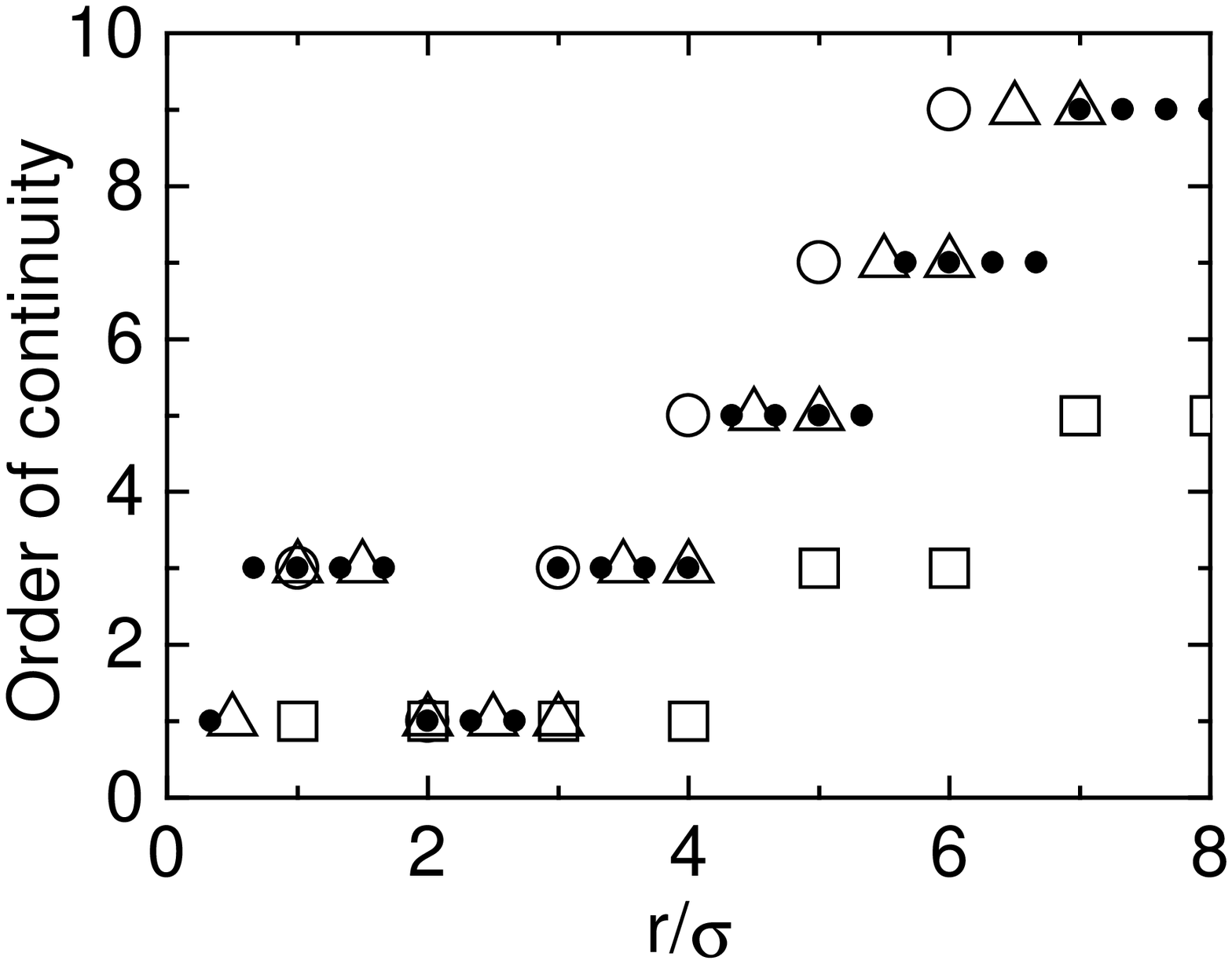,width=4in,clip=}
\caption{Order of continuity of $y(r)$ versus $r/\sigma$ for the SW fluid
with $\lambda=2$ (squares), $\lambda=3/2$ (triangles), and $\lambda=4/3$
(dots), and the HS fluid (circles). The order of continuity is
assumed to be infinity if not plotted.\label{fig2}} 
\end{center}
\end{figure}
where the singularities induced by the hard core are
propagated to. Conjectures about what will occur with arbitrary values of 
$\lambda$ are discussed in the next section.
\section{CONCLUDING REMARKS}
\label{sect_5}
In this paper the analytic properties of the structure function of
square-well (SW) fluids have been studied. The SW potential is a
simple but useful model of more realistic interactions as it includes
not only the volume exclusion effect of the hard core but also a finite
range attractive part. It has been used as a model for fluids,
systems of colloidal particles, microemulsions and micelles
\cite{JSH,CRE,CRO,CGDK,MGHD}. Throughout this paper, it has been assumed 
that the structure function $\psi(r)=r y(r)$ satisfies the well known 
nonlinear Percus-Yevick (PY) integral equation. Despite the apparent 
simplicity of the SW potential, no exact solution of this 
equation has ever been found. The reduction of the PY equation to
a system of nonlinear integral equations unveils the degree of difficulty
of the problem. The structure function, $\psi(r)$, must be defined by
analytical pieces and this is unsuitable for a purely local treatment, except
in the hard-sphere \cite{PY} and sticky-hard-sphere \cite{BX} models, where an 
expression for the Laplace transform of $r g(r)$ was found.\\
Although we have restricted ourselves to the cases $\lambda=2$, $\lambda=3/2$
and $\lambda=4/3$, the general pattern for any value $\lambda$ seems clear.
If $\lambda$ is an integer we have discontinuities only
at $r/\sigma=1,2,\ldots$, the ``resonance'' condition is fulfilled and 
the order of continuity grows as $r$ for large $r$. If $\lambda$ is not
an integer we must expect the appearance of discontinuities at those points
which can be expressed as a linear combination of the fundamental scale
lengths $\sigma$ (the hard core diameter) and $(\lambda-1)\sigma$ (the
SW width), $n \sigma + m (\lambda-1) \sigma$, where $n$ and $m$ are 
integers. If $\lambda$ is a rational number there is some point that
can be obtained by using two different pairs of integers $(n,m)$ and the
location of the discontinuity points repeats periodically after that. That
is not the case for irrational values of $\lambda$. A direct consequence
of the choice of a width well $(\lambda-1)\sigma$ incommensurate with
the hard core diameter $\sigma$ is that the order of continuity at
$r/\sigma=1,2,\ldots$ will possibly coincide with the corresponding to
the HS fluid because the two discontinuities of the interaction potential
do not interfere in this case. In the cases we have discussed ($\lambda=3/2$
and $\lambda=4/3$) there is still some kind of ``resonance'' and this
gives rise to an order of continuity lower in two units to that of
the HS fluid at $r/\sigma=4,5\ldots$ (See Fig. \ref{fig2}). However, the
conjecture expressed in Eqs. (\ref{8a})--(\ref{8e}) for $1 < \lambda < 2$ is
sufficient for the purpose of the search of approximate structure
functions as only this interval is considered physically meaningful in
the applications of the model to simple fluids or colloids. We must also
emphasize that these results have been derived in the context of the
PY equation but, referring only to the key features of the
structure functions, it is possible that they are more general. An
analysis of other integral equations for fluids and the Mayer diagrams not taking
into account by them would be necessary in order to clarify this point.\\
This work was already started in the early days of the theory of liquids. Percus
showed \cite{Percus}, in the context of the PY equation for hard spheres, that
analytic breakings on the structure functions should appear whenever distances
between particle centers along a chain are modified so to accommodate a new
particle. This result explains the discontinuites on the cavity function
derivatives at $r=\sigma$, $2\sigma$, $3\sigma$, \ldots that are listed in
the Equations (\ref{id})--(\ref{ie}). A combination of formal and heuristic 
geometric
arguments were used later by Stillinger \cite{STI} in order to identify the
diagrams which cause discontinuities in the derivatives of $g(r)$ for
hard spheres up to fourth order in the density. The most simple diagrams
are the chains but we have also double chains and triply-connected diagrams
that are not included in the PY approximation. The topological change associated
with the separation or the approaching of a given pair of spheres in the cluster
corresponding to the diagram within a distance which avoids the contact of all the
particles was found to be the origin of a singularity of the structure function. In
that way, the lowest-order double chain gives rise to a singularity
at $r= \sqrt{3} \sigma$ and the lowest-order triply-connected cluster is
responsible for a discontinuity
at $r=\sqrt{8/3} \sigma$ \cite{PG}. Careful Monte Carlo simulations carried
out by Seaton and Glandt \cite{SG} for the fluid of adhesive spheres have
confirmed the existence of discontinuities of the radial distribution
function at $r=\sqrt{8/3} \sigma$, $\sqrt{3} \sigma$ and $2 \sigma$. The
latter is the most striking one and this is taking into account by the
PY solution because it is originated by the chain diagram with two bonds. The
more complex spatial diagrams are responsible for the discontinuities at
$r=\sqrt{8/3} \sigma$ and $\sqrt{3} \sigma$ which are not included in
the PY solution for the sticky hard sphere potential \cite{BX}. Nevertheless,
these two are less prominent features of $g(r)$ than the discontinuity at
$r=2\sigma$ and the PY solution gives a good overall agreement with
Monte Carlo simulation results \cite{SG}. Stillinger \cite{STI} has
even suggested that the full set of singularities of $g(r)$ for the
hard sphere or hard disk case is dense throughout the entire range
$0 \le r < \infty$ as a consequence of the formation or breaking of bonds
on random packings when $r$ (the distance between two given particles in
the cluster) is slightly varied.\\
Analytical approximations
proposed by different routes are equally interesting alternatives to
the solution of the integral equations. These approximations are usually
based on the imposition of continuity conditions at the hard core and the
SW border. For example, the approximation of Nezbeda \cite{Nezbeda}
for a very thin SW interaction is based upon the continuity of
the first and the second derivatives of $y(r)$ at $r=\sigma$. On the other
hand, Yuste and Santos\cite{YusSan} proposed an approximation on the basis
of some physical conditions, among which the continuity of $y(r)$ at
$r=\lambda\sigma$. The latter authors are forced to fix one of the
parameters of their model at its low density value in order to close
the system of nonlinear equations that those parameters satisfy. The
results derived in Secs. \ref{sect_3} and \ref{sect_4} allow the
imposition of continuity conditions on the first derivative of $y(r)$
at $r=\lambda$ and this would close the system of equations of
Yuste and Santos's model in a more convincing way. Similarly, Nezbeda
approximation could possibly be improved by imposing the
continuity of the third derivative at $r=\sigma$. Work
along this line is currently in progress and will be published elsewhere.
\acknowledgments
The author gratefully acknowledges Andr\'es Santos for useful
discussions and a critical reading of the manuscript. Partial support
from the DGICYT (Spain) through Grant no. PB97-1501 and from the
Junta de Extremadura (Fondo Social Europeo) through Grant no. IPR98C019
are also acknowledged.
\appendix
\section{THE HARD-SPHERE FLUID}
In this case the structure function $\psi(t)$ is defined in the
piecewise form given in (\ref{PSIPL}) and the set of equations satisfied
by the analytic functions $\psi_n(t)$, $n=0,1,\ldots$ is obtained
from (\ref{1a})--(\ref{1d}) if the limit $\alpha \rightarrow 0$ is
taken (which is equivalent to take $\epsilon \rightarrow 0$ or
$T \rightarrow \infty$). The derivatives of order $k$, $k \ge 2$ has
already been given in Eqs. (\ref{2a})--(\ref{2d}). By setting $\alpha=0$
in these equations we get
\begin{eqnarray}
\label{ia}
\psi_0^{(k)}(t)&=&-12 \eta \sum_{j=0}^{k-2}\, (-1)^j
\psi_0^{(j)}(1-t) \psi_1^{(k-2-j)}(1)+\left[\mbox{Int}\right]  \\
\noalign{\smallskip}
\label{ib}
\psi_1^{(k)}(t)&=& 12 \eta \sum_{j=0}^{k-2}\, \left\{ (-1)^j 
\psi_0^{(j)}(2-t) \Delta_1^{(k-2-j)}+\psi_0^{(j)}(t-1) \psi_1^{(k-2-j)}(1)
\right\}+\left[\mbox{Int}\right]  \\
\noalign{\smallskip}
\label{ic}
\psi_n^{(k)}(t)&=&12 \eta \sum_{j=0}^{k-2} \left\{ (-1)^j 
\psi_0^{(j)}(n+1-t) \Delta_n^{(k-2-j)}-\psi_0^{(j)}(t-n) 
\Delta_{n-1}^{(k-2-j)}\right\}+\left[ \mbox{Int} \right]\, , \quad n \ge 2 
\end{eqnarray}
where the terms $\left[ \mbox{Int} \right]$ again denote the sum
of several integrals over the functions $\psi_0(t)$, $\psi_1(t)$, \ldots,
whose explicit expressions are not required as these terms are always
continuous at the points of interest $t=1,2,\ldots$. The derivative
jumps $\Delta_n^{(k)}$, $n=0,1,\ldots$, $k=0,1,\ldots$ have been
already defined in (\ref{defi}) and the nonzero symbols $\Delta_n^{(k)}$
corresponding to the lowest value of $k$ for every $n$ can be derived
from (\ref{ia})--(\ref{ic}). The results are
\begin{eqnarray}
\label{id}
\Delta_0^{(4)}&=&-12 \eta \psi_1(1) \Delta_1^{(2)}-24 \eta
\psi_1(1) \psi_0^{(2)}(0)=(12 \eta)^2 \left[\psi(1)\right]^3 \\
\noalign{\smallskip}
\label{ie}
\Delta_n^{(2 n)}&=& (12 \eta)^n \left[-\psi(1) \right]^{n+1}\; , \quad
n \ge 1
\end{eqnarray}
where $\psi(1)=(1+\eta/2)/(1-\eta)^2$ is the PY exact result \cite{PY}
for the structure function at the hard core contact point. Note that the
HS fluid can also be seen as a SW fluid in the
limit $\lambda \rightarrow 1$.


\begin{references}
\bibitem{MTY} G. A. Martynov, {\em Fundamental Theory of Liquids}, (Adam Hilger,
Bristol, 1992).
\bibitem{Rev} J. A. Barker and D. Henderson, {\em Rev. Mod. Phys.}
{\bf 48}, 587 (1976).
\bibitem{PY} J. Percus and G. Yevick, {\em Phys. Rev.} {\bf 110}, 1 (1958).
\bibitem{Percus} J. Percus, {\em The Pair Distribution Function in Classical
Statistical Mechanics} in {\em The Equilibrium Theory of Classical
Fluids}, edited by H. L. Frisch and J. L. Lebowitz (Benjamin, New York, 1964), pp. II-33, II-170.
\bibitem{WT} M. S. Wertheim, {\em Phys. Rev. Lett.} {\bf 10}, 321 (1963); E.
Thiele, {\em J. Chem. Phys.} {\bf 39}, 474 (1963).
\bibitem{BX} R. J. Baxter, {\em J. Chem. Phys.} {\bf 49}, 2770 (1968).
\bibitem{Sharma} P. V. Sharma and K. C. Sharma, {\em Physica A} {\bf 89}, 203
(1977).
\bibitem{Nezbeda} I. Nezbeda, {\em Czech. J. Phys. B} {\bf 27}, 247 (1977).
\bibitem{YusSan} S. B. Yuste and A. Santos, {\em J. Chem. Phys.} {\bf 
101}, 2355 (1994).
\bibitem{Henderson} D. Henderson, W. G. Madden, and D. D. Fitts, {\em J.
Chem. Phys.} {\bf 64}, 5062 (1976).
\bibitem{JSH} J. S. Huang, S. A. Safran, M. W. Kim, G. S. Grest, M.
Kotlarchyk, and N. Quirke, {\em Phys. Rev. Lett.} {\bf 53}, 592 (1984).
\bibitem{CRE} C. Regnaut and J. C. Ravey, {\em J. Chem. Phys.} {\bf 91},
1211 (1989); {\bf 92}, 3250 (1990).
\bibitem{CRO} C. Robertus, J. G. H. Joosten, and Y. K. Levine, {\em Phys.
Rev. A} {\bf 42}, 4820 (1990).
\bibitem{CGDK} C. G. de Kruif, P. W. Rouw, W. J. Briels, M. H. G. Duits,
A. Vrij, and R. May, {\em Langmuir} {\bf 5}, 422 (1989).
\bibitem{MGHD} M. H. G. Duits, R. P. May, A. Vrij, and C. G. de Kruif,
{\em Langmuir} {\bf 7}, 62 (1991).
\bibitem{STI} F. H. Stillinger, {\em J. Comp. Phys.} {\bf 7}, 367 (1971).
\bibitem{PG} A. J. Post and E. D. Glandt, {\em J. Chem. Phys.} {\bf 84}, 4585
(1986).
\bibitem{SG} N. A. Seaton and E. D. Glandt, {\em J. Chem. Phys.} {\bf 87}, 1785
(1987).
\end{references}
\end{document}